# Variational method to study vortex matter in mesoscopic superconductors

W. V. Pogosov [a,*], A. L. Rakhmanov [b], E. A. Shapoval [c]

[a] *Moscow Institute of Physics and Technology, Institutskii per. 9, Dolgoprudny, Moscow region, 141700 Russian Federation*

[b] *Institute for Theoretical and Applied Electrodynamics, Russian Academy of Sciences, Izhorskaya ul. 13/19, Moscow, 127412 Russian Federation*

[c] *All-Russia Scientific-Research Institute of the Metrological Service, Moscow, 119361 Russian Federation*

**Abstract**

A simple variational model is proposed to analyze the superconducting state in long cylindrical type-II superconductor placed in the external magnetic field. In the framework of this model, it is possible to solve the Ginzburg-Landau equations for the states with axially symmetric distributions of the order parameter. Phase transitions between different superconducting states are studied in the presence of external magnetic field and an equilibrium phase diagram of thin cylinder is obtained. The lower critical field of the cylindrical type-II superconductor with arbitrary values of radius and Ginzburg-Landau parameter is found. The field dependence of the magnetization of thin cylinder, which can carry several magnetic flux quanta, is calculated.

## 1. Introduction

Recently, much attention was attracted to the analysis of superconducting state in mesoscopic samples, whose sizes are comparable to the coherence length $x(T)$ or London penetration depth $l(T)$ [1-15]. Numerous experimental and theoretical works are devoted to the study of the magnetic response of different shaped mesoscopic samples. It is known that below the first critical field $H_{c1}(T)$ there are no vortices inside a type-II superconductor (the Meissner state). A multi-vortex state (the Abrikosov phase) becomes energetically more favorable between $H_{c1}(T)$ and bulk upper critical field $H_{c2}(T)$. In the vicinity of $H_{c2}(T)$ a phase with the giant-vortex in the center of the sample carrying $L > 1$ flux quanta has the lowest energy, and Abrikosov vortices merge into this structure [5-14]. In this case, the order parameter is strongly suppressed in the inner part of the sample, and this state can be referred to a surface superconductivity. With further increasing of the applied magnetic field $H$, phase transitions between different giant-vortex states are observed. The superconductor comes to the normal state at $H = H_{c3}(T)$ (where $H_{c3}(T)$ is the surface or the third critical field). Note that both the values of $H_{c1}(T)$ and $H_{c3}(T)$ depend on the sample sizes and geometry.

In this paper, we apply a variational method for the study of the vortex state in a long type-II superconducting cylinder. Based on the trial function for the order parameter, it is possible to find the approximate solution to the Ginzburg-Landau (GL) equations for the Meissner state, single-vortex state, and giant-vortex state. The model enables us to find explicit expressions for the dependence of the lower critical field of the cylinder $H_{c1}^{(cyl)}$ on its radius $R$ and GL parameter $k$. Phase boundaries between different superconducting states are

---

calculated and an equilibrium phase diagram of thin cylinder is obtained. We found that states with axial symmetric distribution of the order parameter are energetically more favorable with respect to the Abrikosov phase if the radius of the cylinder is smaller than $2.75x(T)$. In this case, equilibrium magnetic properties of sample can be described in the framework of the variational model. We calculate the magnetization curve of thin cylinder, which can carry only several quanta of magnetic flux.

Our results for the lower critical field are in a good agreement with the numerical solutions of GL equations [8], where the superconductors with $k \gg 1$ were studied. In the case $k \gg 1$, it is possible to neglect the spatial distribution of the magnetic field inside the cylinder when calculating the lower critical field at $R \sim x(T)$ and to reduce the system of GL equations to a single one. Within the framework of our variational model, the spatial distribution of the magnetic field for the Meissner state and single-vortex state can be taken into consideration accurately, which is important at $k \sim 1$.

## 2. Ginzburg-Landau theory for the superconducting state in the cylinder

Consider a cylindrical type-II superconductor in the uniform applied magnetic field $H$ parallel to the cylinder axis. The sample is much longer than $l(T)$, so both the order parameter and the magnetic field are constant along its axis. Let us introduce the dimensionless quantities: all the distances are measured in units of $x(T)$, the magnetic field is measured in units of bulk upper critical field $H_{c2}(T)$, and the order parameter is measured in units of $\sqrt{-a/b}$ with $a$, $b$ being the GL coefficients. In this paper, we study the states with axially symmetric distributions of the order parameter and the magnetic field inside the cylinder. In this case, the order parameter can be presented as $f(r)\exp(ijL)$ (the so called "$L$-state"), where $r$ is the dimensionless radius-vector in the cross-sectional area of the cylinder, $j$ is the azimuthal angle, and $L$ is the angular quantum momentum ($L= 0, 1, 2,\ldots$). If $L= 0$ this state is vortex-free or the Meissner state, if $L= 1$ it is the single-vortex state, and at $L > 1$ it is the giant-vortex state.

The Gibbs free energy $G$ of the cylinder can be written in the form of the GL functional:

$$G = \frac{2p}{k^2} \int_0^r rdr \left[ f'^2 + \frac{1}{2}\left(1 - f^2\right)^2 + f^2\left(a - \frac{L}{r}\right)^2 + k^2 b^2 - 2k^2 hb \right], \qquad (1)$$

where $r$, $a$, $b$ are dimensionless radius of the cylinder, vector potential, local magnetic field respectively, and $h$ is external field. The system of GL equations and the relationship between $a$ and $b$ are given by:

$$-\frac{1}{r}\frac{d}{dr}\left(r\frac{df}{dr}\right) + f^3 - f + f\left(a - \frac{L}{r}\right)^2 = 0, \qquad (2)$$

$$\frac{dh}{dr} = \frac{f^2}{k^2}\left(a - \frac{L}{r}\right), \qquad b = \frac{1}{r}\frac{d(ra)}{dr}. \qquad (3)$$

The boundary conditions for the magnetic field and the order parameter are:
$$b'(0) = 0, \ b(r) = h; \qquad (4)$$
$$f'(r) = 0; \qquad (5)$$
$$rf(r)^{-2}b'(r) = -L, \ r \to 0. \qquad (6)$$

Note, that here we use the standard in GL theory boundary condition (5) for the order parameter on the boundary of superconductor and vacuum [16, 17]. Condition (6) can be found from Eqs. (3) [17]. As it follows from GL equations, their solutions behave as $f(r) \sim r^L$ at $r \to 0$ [17].

The total Gibbs free energy (1) of the cylinder is a sum of two terms $G_{op}$ and $G_m$, related to the spatial variation of the order parameter (the sum of first two terms in (1)) and to the energy of the magnetic field and the supercurrents (the sum of last three terms in (1)). Using (3), (4) and (6), it is easy to find:

$$G_m = -2p h r k^2 f^{-2}(r) b'(r) + 2p L k^2 (b(0) - 2h). \tag{7}$$

It is a rather complicated task to solve the system of coupled equations (2) and (3) due to their non-linearity. In the next section we shall describe an approximate method of solution to system (2), (3).

## 3. Variational ansatz

Instead of solving of the GL equations, it is possible to use trial functions for the local coordinate dependence of the order parameter. We propose the following trial function for the spatial distribution of the order parameter:

$$f = \begin{cases} f_L \left(\dfrac{r}{r_L}\right)^L \left(2 - \dfrac{r^2}{r_L^2}\right)^{L/2}, & r \leq r_L; \\ f_L, & r \geq r_L, \end{cases} \tag{8}$$

with $f_L$ and $r_L$ being the variational parameters. The values of $f_L$ and $r_L$ can be calculated self-consistently by minimization of the free energy. However, at $L = 1$ this leads to cumbersome expressions. So we use another trial function, which was proposed in Ref. [19]:

$$f(r) = \frac{hr}{\sqrt{r^2 + x_v^2}}, \tag{9}$$

where $h$ and $x_v$ are the variational parameters. The application of this function leads to more accurate value of the bulk lower critical field $h_{c1}^{(b)}$ [18,19]. But trial function (9) does not meet boundary condition (5) since its derivative can not be equal to zero on the surface of the cylinder. However, $f'(r)$ is small even at $r \sim 1$ ($R \sim x(T)$). In addition, we found that both models give very similar results. Thus, below for the sake of simplicity we shall use trial function (9) for single - vortex state ($L = 1$), while for giant - vortex states ($L > 1$) and vortex - free state ($L = 0$) trial function (8) is used. An accuracy of our variational model is discussed in Sections 4 and 5.

### 3.1. Meissner state

Let us find the energy of the vortex - free state (Meissner state). The value of the order parameter $f_0$ (8) is assumed to be constant throughout the sample. In this case the local magnetic field can be easily found from the second GL equation (3) and the boundary conditions (5):

$$b(r) = h \frac{I_0(f_0 r/k)}{I_0(f_0 r/k)}, \tag{10}$$

where $I_n$ (and further on $K_n$) are the Bessel functions of imaginary argument. The total Gibbs free energy of the cylinder in the vortex-free state in this approximation is:

$$G = -2p\, h^2\, r\, f_0^{-1} \frac{I_1(f_0 r/k)}{I_0(f_0 r/k)} + \frac{p\, r^2}{2k^2}\left(1 - f_0^2\right)^2. \tag{11}$$

The value of $f_0$ at arbitrary $h$ can be found calculated by minimization of $G$. Thus, we can find the approximate solution to the system of GL equations for the vortex-free state.

*3.2. Single-vortex state*

The local order parameter is described by function (9). In this case, the local magnetic field at $L = 1$ can be found from the Eqs. (3) when the order parameter is given by Eq. (9):

$$b(r) = VK_0\left(hk^{-1}\sqrt{r^2 + x_v^2}\right) + tI_0\left(hk^{-1}\sqrt{r^2 + x_v^2}\right), \qquad (12)$$

where $z$ and $t$ are constants. One can find the values of $z$ and $t$ from the boundary conditions (4) and (6):

$$V = \Omega_{01}^{-1}\left[hI_1(hx_v/k) + \frac{hk}{x_v}I_0\left(hk^{-1}\sqrt{r^2 + x_v^2}\right)\right], \qquad (13)$$

$$t = \Omega_{01}^{-1}\left[hK_1(hx_v/k) - \frac{hk}{x_v}K_0\left(hk^{-1}\sqrt{r^2 + x_v^2}\right)\right], \qquad (14)$$

where we define the matrix elements $\Omega_{st}$ as

$$\Omega_{st} = \left[K_s\left(hk^{-1}\sqrt{r^2 + x_v^2}\right)I_t(hx_v/k) + (-1)^{s+t+1}I_s\left(hk^{-1}\sqrt{r^2 + x_v^2}\right)K_t(hx_v/k)\right], \; s, t=0,1.$$

Taking into account Eqs. (7), (12)-(14) we obtain the following expression for the magnetic free energy of the cylinder:

$$G_m = Ah^2 + Bh + C, \qquad (15)$$

where

$$A = \frac{2p\sqrt{r^2 + x_v^2}\,k\,\Omega_{11}}{h\,\Omega_{01}}, \quad B = \frac{4pk^3}{hx_v\,\Omega_{01}} - 4pk^2, \quad C = -2p\frac{h\,\Omega_{00}k^3}{x_v\Omega_{01}}.$$

Using Eqs. (1) and (11) it is possible to find the energy related to the spatial variation of the order parameter:

$$G_{op} = \frac{pr^2}{k^2}\left[\frac{1}{2}(1-h^2)^2 + \frac{h^4}{2} + \frac{h^2 x_v^2(1-h^2)}{r^2}\ln\left(1 + \frac{r^2}{x_v^2}\right) - \frac{h^4 r^2}{2(r^2 + x_v^2)} + \frac{h^2(r^2 + 2x_v^2)}{r^2(r^2 + x_v^2)}\right]. \qquad (16)$$

So we derived the dependence of the total Gibbs free energy of the cylinder $G = G_m + G_{op}$ on the applied magnetic field $h$ and variational parameters $h$, $x_v$. At arbitrary $h$, the values of $h$ and $x_v$ can be calculated by minimization of the Gibbs free energy $G$ with respect to $h$ and $x_v$. Thus, in the framework of the variational model it is possible to find the approximate solutions to the GL equations (2), (3) at $L = 1$.

*3.3. Giant-vortex states*

In this subsection, we shall describe an approximate method to solve equations (2) and (3) at $L > 1$. The local order parameter is given by Eq. (8). When the giant vortex is located in the center of thin cylinder at $h > 1$ ($H > H_{c2}(T)$), the order parameter has a ring-like distribution. In this case the magnetic field undergoes only slight spatial variation in the superconductor, therefore the contribution of the spatial variation of field to the free energy can be neglected [20, 21]. If the magnetic field is constant inside the cylinder, the vector potential is $a = hr/2$. So it is easy to find the total Gibbs free energy of the system $G$ from Eq. (1) by a straightforward integration. However, the final results for the free energy and the local magnetic field (which we shall find below) are cumbersome and we do not write it here. Similar to the previous cases, the values of variational parameters can be calculated by minimization of free energy $G$. In this case, the expression for the free energy is biquadratic with respect to $f_L$, which makes the minimization procedure much simpler. Thus, the spatial distribution of the order parameter and the free energy of the cylinder can be found.

The local magnetic field can be calculated from Eqs. (3), (4), and (8), using known local order parameter:

$$b(r) = h - \frac{1}{k^2} \int_r^r f(x)^2 \left(\frac{hx}{2} - \frac{L}{x}\right) dx. \tag{17}$$

The giant-vortex phase exists if $h > 1$ ($H > H_{c2}(T)$) and it will be shown in Section 5 that under this condition the error in magnetization arising due to neglecting of the spatial variation of field is just a few percent even at $L = 1$ and small $k$.

## 4. The critical fields and phase diagram of the cylinder

The lower critical field of the cylinder $H_{c1}^{(cyl)}$ is the value of external magnetic field $h$ at which the energies of Meissner state ($L = 0$, Eq.(11)) and single-vortex state ($L = 1$, Eqs. (15) and (16)) become equal. Equating these energies, it is possible to obtain the dependence of dimensionless lower critical field $h_{c1}^{(cyl)}(r)$ on the variational parameters:

$$h_{c1}^{(cyl)}(r) = -\frac{1}{2}\left(A + 2pr\, f_0^{-1} I_1(f_0 r/k) I_0^{-1}(f_0 r/k)\right)^{-1} \times$$

$$\times \left\{-B - \left[B^2 - 4\left(A + 4pr\, f_0^{-1} I_1(f_0 r/k) I_0^{-1}(f_0 r/k)\right)\left(C + G_{op} - \frac{pr^2}{2k^2}(1 - f_0^2)^2\right)\right]^{1/2}\right\}. \tag{18}$$

We found by numerical minimization of the total Gibbs free energy that the variational parameters at the transition from $L = 0$ to $L = 1$ state can be approximated by the following explicit expressions:

$$f_0 = \left[1 - \left(\frac{1}{0.565 r^2}\right)^2\right]^{0.8}, \tag{19}$$

$$h = \left[1 - \left(\frac{1}{0.56 r^2}\right)^2\right]^{0.6}, \quad x_v = x_v^{(b)}\left[1 - \left(\frac{1}{10 r^2}\right)^{0.75}\right]^5, \tag{20}$$

where $x_v^{(b)}$ is the value of $x_v$ at $r \to \infty$. It can be found from the condition $\partial G / \partial x_v^{(b)} = 0$ at $r \to \infty$:

$$x_v^{(b)} - \sqrt{2}\left[1 - \frac{K_0^2(x_v^{(b)}/k)}{K_1^2(x_v^{(b)}/k)}\right] = 0. \tag{21}$$

At $\kappa \gg 1$ we have: $x_v^{(b)} \approx \sqrt{2}$. Eqs. (18)-(21) give the solution for $h_{c1}^{(cyl)}(r)$.

When the radius of the cylinder $r \gg 1$ ($R \gg x(T)$) the above expression for $h_{c1}^{(cyl)}(r)$ reduces to formula, which was obtained in the framework of London approximation in Ref. [20]:

$$h_{c1}^{(cyl)}(r) = \left(h_{c1}^{(b)} - \frac{K_0(r/k)}{2k^2 I_0(r/k)}\right)\frac{I_0(r/k)}{I_0(r/k) - 1}.$$

The dependence $h_{c1}^{(cyl)}(r)$ (18) is shown in Fig. 1 for $k = 5$ (curve 1). In Fig. 2 we plotted this dependence at different $k$ values. In Fig 2 the lower critical field is also shown at $\kappa \to \infty$ (dot line), which was calculated by numerical solution to the GL equations in Ref. [10]. The deviation of variational result from the numerical one does not exceed few percent. In addition, we found that dimensionless $h_{c1}^{(cyl)}(r)$ behaves in a similar way (scaling behavior) for different $k$ starting with $k \approx 3$ (see Fig. 2).

The dimensionless third critical field of the cylinder $h_{c3}^{(cyl)}(r)$ is found from the linearized GL equation (2) [8,10], but $h_{c3}^{(cyl)}(r)$ can also be calculated in the framework of present variational model with good accuracy if the cylinder is not thicker than several $x(T)$. In contrast to the linear approach, our approximation allows finding phase boundaries between different $L$-states and the magnetization. The value of $h_{c3}^{(cyl)}(r)$ can be defined from the condition of equality of the energy of $L$-state (which is always energetically more favorable than the Abrikosov phase above $h = 1$ ($H = H_{c2}(T)$)) and normal state. At each $r$ we should chose the maximum value of $h_{c3}^{(cyl)}(r)$ among those, corresponding to different $L$. The resulting dependence $h_{c3}^{(cyl)}(r)$ is shown in Fig. 1 (curve 2). The numbers below the curve 2 characterize the angular quantum momentum $L$ of corresponding states. The difference between the exact values of $h_{c3}^{(cyl)}(r)$, found from the linearized GL equation (dot line), and variationally-calculated $h_{c3}^{(cyl)}(r)$ is few percent at $L < 5$, which indicates good accuracy of variational model. At $L > 5$ the deviation of variational $h_{c3}^{(cyl)}(r)$ from the exact result increases.

An equilibrium phase diagram (Fig. 1) includes the vortex-free state (region below curve 1), the vortex state (between curves 1 and 2) and the normal state (upper than curve 2). Phase boundaries between different $L$-states can be calculated using the results of Section 3, where the energies of these states were found (Fig. 1). However, the energy of superconductor in the Abrikosov phase, which is energetically more favorable below $h = 1$ ($H = H_{c2}(T)$), can not be calculated in the framework of our model, because the local order parameter is not axially symmetric in this case. Nevertheless, as can be seen from the phase diagram (Fig. 1), $L$-states have the lower energy as compared to the Abrikosov phase at small cylinder radius $R \leq 2.75x$, since the field of transition from $L = 1$ to $L = 2$ state is bigger than $h = 1$ ($H = H_{c2}(T)$) in this case. Therefore equilibrium magnetic properties of such a sample can be studied in the framework of present variational model (see the next section). Note in conclusion of the section that various phase diagrams for different shaped samples were studied in Ref. [8-13].

## 5. Equilibrium magnetization of thin cylinder

The magnetization of a mesoscopic superconductor differs crucially from that for a bulk superconductor. It follows a set of curves, corresponding to different number of vortices and different $L$-states inside [6, 7-14, 21, 22]. Let us find the equilibrium magnetization of thin cylinder with radius smaller than $2.75x(T)$, where the multi-vortex phase is energetically unfavorable, relative to $L$ - states. From the definition, the magnetization is $M = (<b> - h) / 4\pi$ where $<b>$ is the averaged magnetic field over the volume of superconductor. The spatial distributions of the magnetic field and the order parameter inside the cylinder were found in Section 3. The results for magnetization in the cases of different cylinder radiuses at $k = 5$ are shown in Fig. 3. In Fig. 3 (a) the magnetization curve is plotted for the cylinder with $R =$ 1.2$x(T)$, which can not accommodate any vortex. Fig.3 (b) and (c) correspond to the cases $R$ = 2.25$x(T)$ and 2.5$x(T)$, in which the cylinders can accommodate giant vortices with $L = 2$ and $L = 3$ respectively. Jumps in the magnetization correspond to the transitions between different $L$-states.

Let us check the accuracy of the assumption that the spatial vatiation of the magnetic field can be neglected when calculating the total free energy of a giant-vortex state. For this purpose, we compare the magnetization for single-vortex state ($L = 1$), calculated in the framework of approaches of Subsections 3.2 and 3.3. In the former case this contribution is taken into account while in the later case it is neglected. We found that the difference between these results is just a few percent at $h > 1$ ($H > H_{c2}(T)$), which demonstrates a good accuracy of our approach. Note in conclusion that the similar variational procedures were applied in Ref. [18, 23, 24, 25] to study the magnetic properties of a bulk type-II superconductor in the entire range of external magnetic fields between $H_{c1}$ and $H_{c2}$.

## 6. Conclusion

In this paper, we apply a variational method for the analysis of vortex state in mesoscopic cylindrical type-II superconductors placed in the external magnetic field. Instead of solving of the coupled nonlinear Ginzburg-Landau equations, it is possible to use the trial function to describe the spatial distribution of the order parameter in the cylinder. The local magnetic field inside the cylinder can be found explicitly from the second Ginzburg-Landau equation. The model enables us to find an approximate solution to the Ginzburg-Landau equations for different states with axial symmetric distribution of the order parameter. Within the framework of the model, we calculate the dependence of the lower critical field on the radius of the cylinder at arbitrary Ginzburg-Landau parameter $k$ and the equilibrium phase diagram. We find that states with axial symmetric distributions of the order parameter are energetically more favorable with respect to the Abrikosov phase if the cylinder radius is smaller than about 2.75$x$. We calculate the field dependencies of the magnetization of such cylinders, which is able to accommodate only several magnetic flux quanta.

## Acknowledgments

We are grateful to K. I. Kugel, L. G. Mamsurova, and K. S. Pigalskiy for useful discussions. This work is supported by the Russian Foundation for Basic Research (RFBR), grants #00-02-18032 and #00-15-96570, by the joint INTAS-RFBR program, grant #IR-97-1394, and by the Russian State Program 'Fundamental Problems in Condensed Matter Physics'.

**Figure captions**

**Fig. 1.** The equilibrium phase diagram of the cylindrical type-II superconductor ($k = 5$) in the applied magnetic field. The solid line 1 corresponds to the lower critical field $H_{c1}^{(cyl)}$; the solid line 2 corresponds to the surface critical field $H_{c3}^{(cyl)}$ calculated by the variational method. The numbers below curve 2 denote the angular quantum momentum of superconducting states. The dashed line corresponds to the exact surface critical field found from the linearized Ginzburg-Landau equations.

**Fig. 2.** The lower critical field of the cylinder $H_{c1}^{(cyl)}$ vs cylinder radius $R$ at different $k$ values. The solid lines correspond to the variational calculation at $k = 0.8, 1.5, \infty$; the dot line corresponds to the numerical solution to the GL equations [10] at $k = \infty$.

**Fig. 3.** The equilibrium field dependence of the magnetization of the thin cylinder with the radius $R$ at $k = 5$, (a) $R = 1.2x(T)$, (b) $R = 2.25x(T)$, (c) $R = 2.5x(T)$. Jumps in the magnetization correspond to the transitions between different $L$ - states.

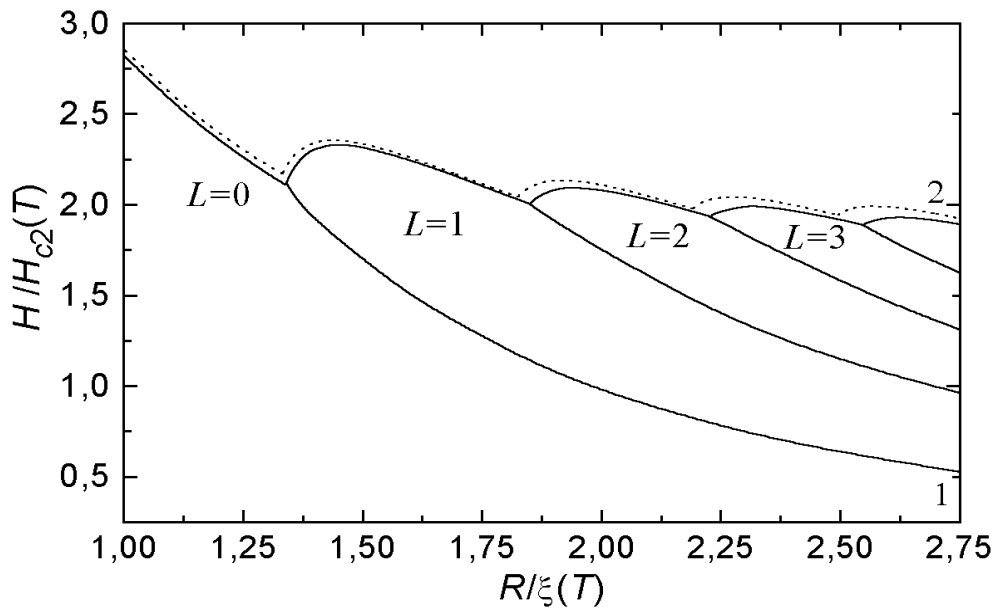

Fig. 1

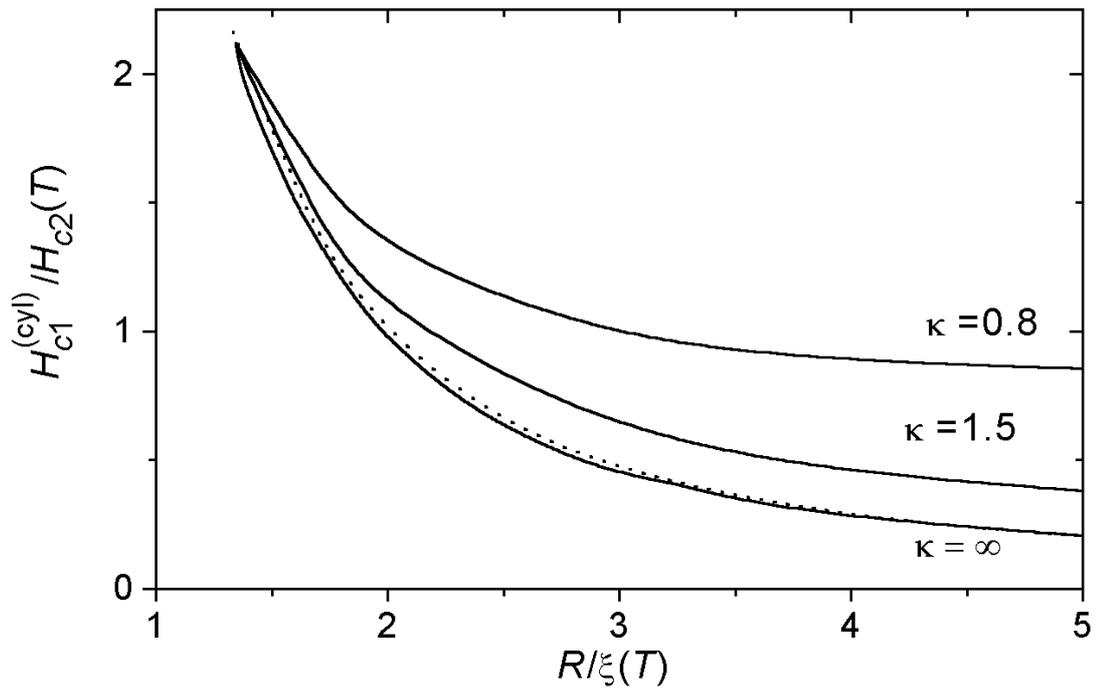

Fig. 2

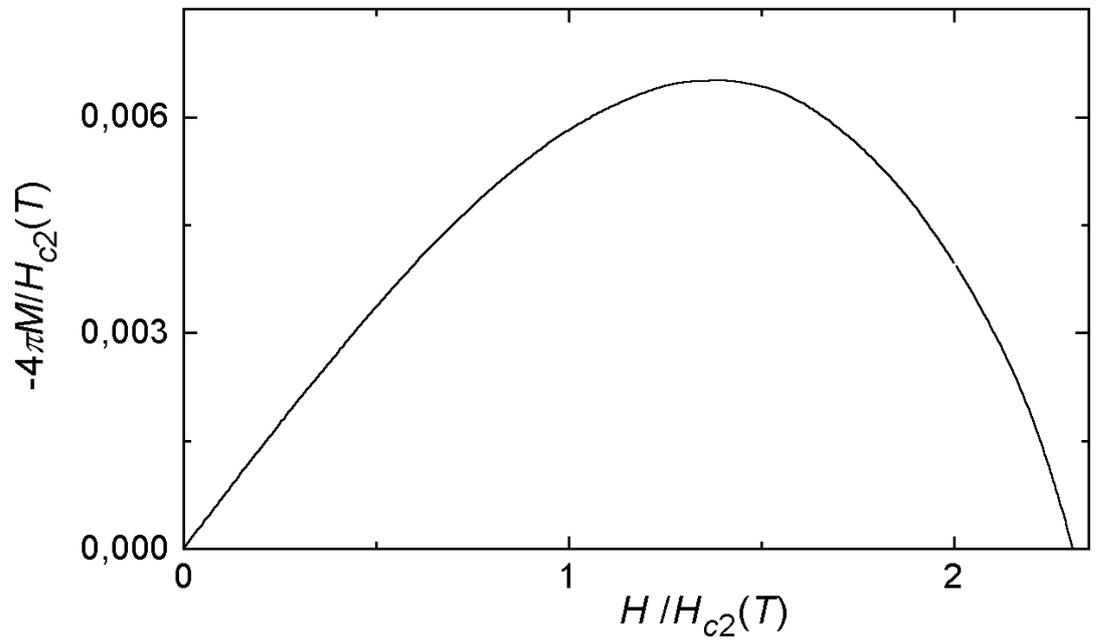

Fig. 3 (a)

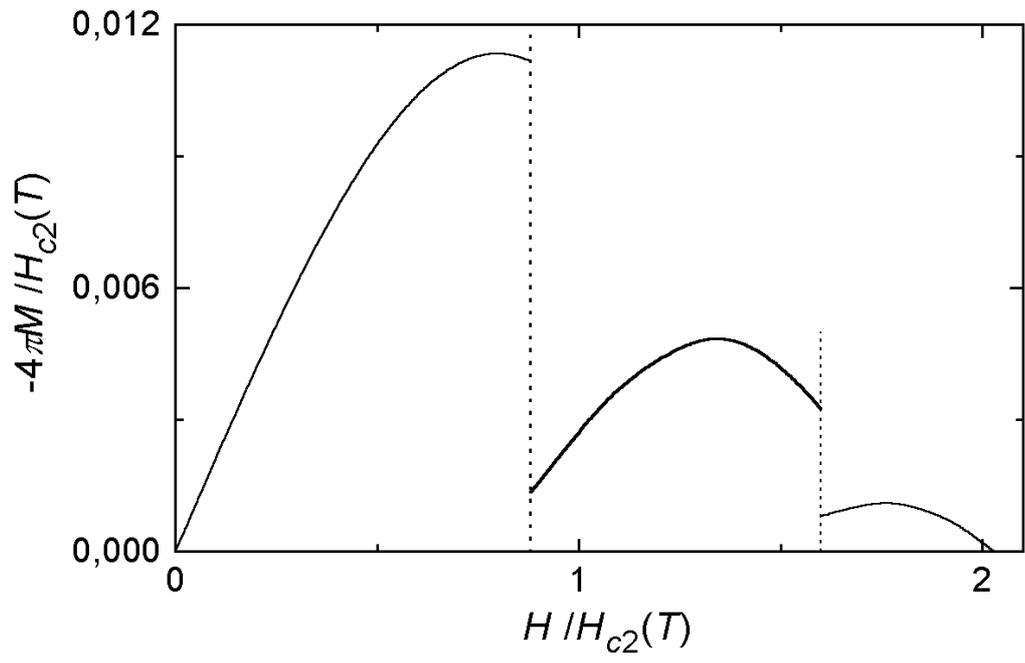

Fig. 3 (b)

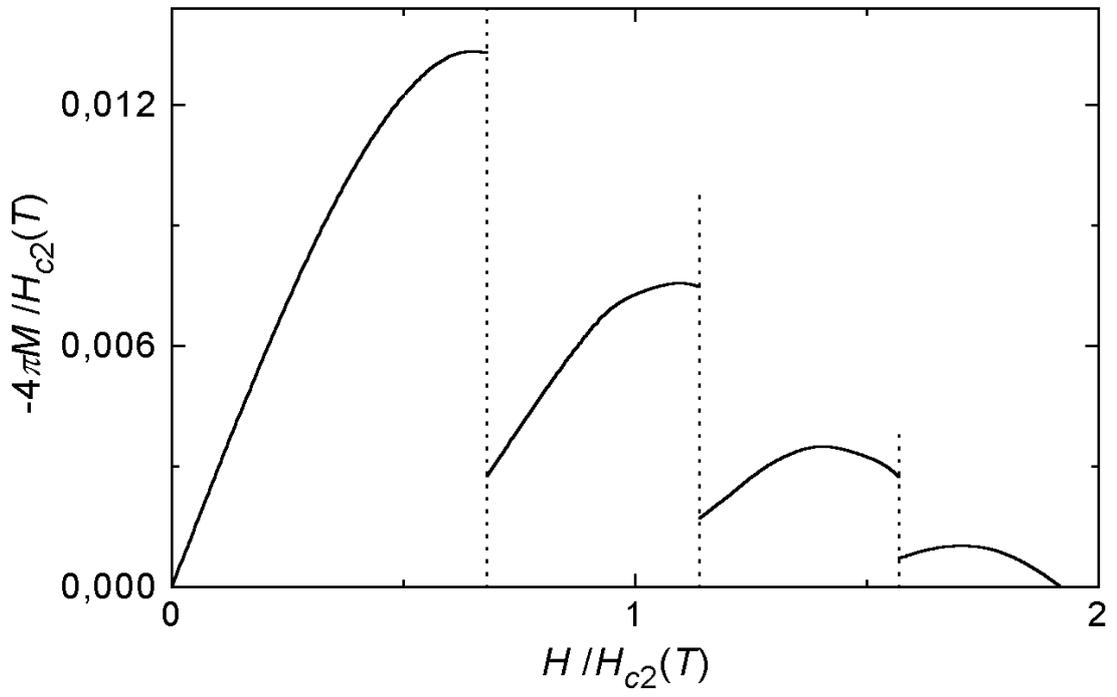

Fig. 3 (c)